\documentclass{emulateapj}

\usepackage{amssymb}
\usepackage{amsmath}
\usepackage{natbib}
\usepackage{txfonts}
\usepackage[colorlinks=true,citecolor=blue,linkcolor=blue]{hyperref}
\usepackage{microtype}
\usepackage{color}
\usepackage{graphicx}
\usepackage{epstopdf}

\def\inpe{1}
\def\capes{2}
\def\towson{3}

\shorttitle{The influence of quantum vacuum friction on pulsars}
\shortauthors{Coelho, Pereira and de Araujo}

\begin{document}

\title{The influence of quantum vacuum friction on pulsars}

\author{Jaziel G. Coelho\altaffilmark{\inpe}, Jonas P.~Pereira\altaffilmark{\capes,\towson} and Jos\'e C. N. de Araujo\altaffilmark{\inpe}}

\altaffiltext{\inpe}{Divis\~ao de Astrof\'isica,        Instituto Nacional de Pesquisas Espaciais, Avenida dos Astronautas 1758,                     12227--010 S\~ao Jos\'e dos Campos,            SP, Brazil}
\altaffiltext{\capes}{CAPES Foundation, Ministry of Education of Brazil, Bras\'ilia, Brazil}

\altaffiltext{\towson}{Department of Physics, Astronomy and Geosciences, Towson University, 8000 York Road, Towson, Maryland 21252-0001, USA}

\begin{abstract}
We firstly revisit the energy loss mechanism known as quantum vacuum friction (QVF), clarifying some of its subtleties. 
Then we investigate the observables that could easily differentiate QVF from the classical magnetic dipole radiation for pulsars with braking indices (n) measured accurately. 
We show this is specially the case for the time evolution of a pulsar's magnetic dipole direction ($\dot{\phi}$) and surface magnetic field ($\dot{B}_0$). As it is well known in the context of the classic magnetic dipole radiation, $n<3$  
would only be possible for positive $(\dot{B}_0/B_0 + \dot{\phi}/\tan\phi)$, which, for instance, leads to $\dot{B}_0>0$ ($\dot{\phi}>0$) when $\phi$ ($B_0$) is constant. On the other hand, we show that QVF can result in very contrasting predictions with respect to the above ones. Finally, even in the case $\dot{B}_0$ in both aforesaid models for a pulsar has the same sign, for a given $\phi$, we show that they give rise to different associated timescales, which could be another way to falsify QVF.
\end{abstract}


\altaffiltext{}{jaziel.coelho@inpe.br; jpereira@towson.edu;  jcarlos.dearaujo@inpe.br}

\maketitle


\section{Introduction}
\label{int}
The crucial concept of the rotational energy of a neutron star (NS) as an energy reservoir for the pulsar's activity, put forward by \cite{1968Natur.218..731G} and \cite{1968Natur.219..145P}, is a manner to explain its kinematical source of energy loss \citep{1969Natur.221..454G}. 
A pulsar's surface magnetic
field has since then been estimated by equating its temporal change of rotational energy  

\begin{equation}\label{eq:Edot}
\dot{E}_{\rm rot} =  I\omega\dot{\omega},
\end{equation}
to the radiating power of a rotating magnetic point-like dipole in vacuum \citep[see e.g.,][]{1975ctf..book.....L,2001thas.book.....P},
\begin{equation}\label{eq:dipole}
P_{\rm dip}=-\frac{2}{3}\frac{m^2_\bot \omega^4}{c^3}
\end{equation}
where $\omega$ and $\dot{\omega}$ are the star's angular velocity and its derivative, respectively, $I$ its moment of inertia, $m_\bot=m_0\sin \phi$ the component of the magnetic dipole $m_0$ perpendicular to the axis of rotation (which is parallel to $\vec{\omega}$), and $\phi$ the angle the magnetic dipole makes with $\vec{\omega}$. 

One can readily show that 
\begin{equation}
m_0= \frac{B_0R^3}{\sqrt{2}}\label{m}
\end{equation}
when $B_0$ is taken as the mean surface magnetic field (coming from a magnetic dipole) of a star of radius $R$. Thus, from Eqs. (\ref{eq:Edot}), (\ref{eq:dipole}) and (\ref{m}) we have
\begin{equation}\label{eq:Bmax}
B_0\sin\phi=\left(\frac{3 c^3}{4 \pi^2} \frac{I}{R^6} P \dot{P} \right)^{1/2}\, ,
\end{equation}
with $P=2\pi/\omega$ and $\dot{P}$ are the rotational period and the spin down rate of a pulsar (observational parameters), while the star's moment of inertia and radius are model dependent parameters.
General considerations on the nature of pulsars have been traditionally obtained in the literature
from the application of the previous formulas for systems with a representative mass $M=1.4 M_\odot$ and
radius $R=10$ km [fiducial parameters \citep[see e.g.,][]{2001thas.book.....P}], implying a moment of inertia of the order of $I\sim 10^{45}$~g.cm$^2$.
For instance, a class of NSs known as high magnetic field pulsars (high-B pulsars) \citep{2015ApJ...799...23B, 2011ApJ...734...44Z, 2011AIPC.1379...60N} would have $B_0$'s higher than the scale field of QED, namely $B_c\doteq m_e^2c^3/(e\hbar)\approx 4.4\times 10^{13}G$ \citep{2010PhR...487....1R}. Ordinary pulsars would have $B_0\lesssim B_c$
\citep[see e.g.,][]{1983bhwd.book.....S, 2000ApJ...541..367C}.

Nevertheless, whenever the magnetic field of a given system is close to $B_c$, quantum effects should play a noteworthy role there. Thus, one would expect that a more accurate description of pulsars, still in the classical point of view, could only be attained with the use of generalizations to the Maxwell Lagrangian, such as the Euler-Heisenberg Lagrangian for QED \citep{2010PhR...487....1R}.

In this regard, it seems that the so-called quantum vacuum friction (QVF) effect, put forward by Dupays and collaborators \citep{2008EL.....8269002D, 2012EL.....9849001D}, has been overlooked in the literature. QVF can be understood as follows. It was shown by Born and Infeld in their seminal work \citep{1934RSPSA.144..425B} that any nonlinear theory of the electromagnetism in vacuum described by a Lagrangian density ${\cal L}$ can be  completely exchanged for the Maxwell theory in a convenient nonlinear medium. This is very important and powerful in the sense that one does not need to derive all involved properties and byproducts of ${\cal L}$, but rather work with Maxwell equations in continuous media. This means that for any ${\cal L}$ the concept of magnetization is present whenever non-null magnetic fields take place and its physical implications are as real as any tangible material medium. For the astrophysical case, it is well known that the magnetic dipole approximation already leads to the correct order of magnitude for the relevant physical quantities there and thus should be the starting point of any model \citep{2001thas.book.....P}. This is exactly where QVF comes into play: the effective magnetized medium (naturally outside the star) should interact with the magnetic dipole of the system (source of the magnetic field), leading it to eventually lose energy. Such an energy loss is due to the torque the magnetic field from the magnetization exerts on the rotating magnetic dipole. We shall show in detail subsequently that such a resultant (time-averaged) torque is anti-parallel to the angular velocity of the star and linearly dependent upon its norm (thus showing that the associated force is dissipative), which slows the star's rotation down while converting the rotational energy into heat. One could thus picture QVF as an energy loss mechanism due to the ``friction'' of the magnetized vacuum with the rotating star, exactly as its name suggests \citep{2008EL.....8269002D}. The medium of the star itself is only important to determine the properties of the magnetic dipole (such as its magnitude and spatial orientation with respect to the axis of symmetry of the system) and does not directly contribute to QVF. 

From the above reasoning, one clearly sees that QVF has an utterly different physical nature than that one underlying the radiation of a rotating magnetic dipole. Therefore, it is meaningless to automatically assume that the former is smaller than the latter, even within the scope of small nonlinear corrections to the classical Maxwellian Lagrangian (known as weak field nonlinear Lagrangians). In this case, what does happen is that the corrections to the classic magnetic dipole radiation due to the nonlinearities of the Lagrangian are very small and thus could be totally disregarded when other types of energy losses are also involved. 

Besides modifying Eq.~(\ref{eq:Bmax}), QVF also modifies the expression for the so-called braking index, with important consequences.
Recall that this quantity is defined as \citep[see e.g.,][]{2001thas.book.....P}
\begin{equation}
n= \frac{\omega\,{\ddot\omega}}{\dot{\omega}^2}\label{n}
\end{equation}
where $\ddot\omega$ is the second time derivative of the angular velocity. It is well known in the literature that when energy losses are only related to the magnetic dipole radiation, $ n = 3$; this fact is in disagreement with observations, which show
that $ n < 3$. We shall see later in this paper that $ n < 3$ is naturally the case whenever QVF also features in the energy loss budget of pulsars, along with the classic magnetic dipole radiation.

It is worth mentioning that there are several scenarios that challenge the magnetic dipole model, like the one involving the accretion of fall-back material via a circumstellar disk \citep{2016MNRAS.455L..87C}, relativistic particle winds \citep{2001ApJ...561L..85X,2003A&A...409..641W}, and modified canonical models to explain the observed braking index ranges \citep[see e.g.,][]{1997ApJ...488..409A,2012ApJ...755...54M}, and references therein for further models). However, no model has been developed yet explaining satisfactory all measured braking indices, nor any of the existing ones has been ruled out by current data. Therefore energy loss mechanisms for pulsars are still under continuous debate.

Our aim in this work is to explore QVF in the context of pulsars
(in particular, those ones that have braking indices measured accurately) solely along with the classic magnetic dipole radiation, since, as we shall show, it already can explain several aspects of their phenomenology. Following this reasoning, we also explore the QVF model to make evolutionary analyses of the pulsars' characteristic parameters, seeking for quantities that could easily contrast it with the classic magnetic dipole radiation and ultimately even falsify QVF.

This paper is organized as follows. In the next section we revisit QVF within compact stars and derive its associated energy loss and rotational period evolution expression for weak field nonlinear Lagrangians, focusing mainly on QED. Section \ref{braking} is devoted to the investigation of the braking indices and the self-consistency of 
the model when both QVF and the classic dipole radiation are responsible for the spin-down of pulsars, for the simpler case in which only the evolution of $P$ is of relevance. In Section \ref{mag_evol} we elaborate upon the evolution of other pulsars' characteristic parameters, such as $B_0$ and $\phi$, in the context of QVF. Finally, in Section \ref{discussion} we discuss the principal issues raised by QVF within the scope of pulsars. We work here with Gaussian units.

\section{QVF in stars revisited}
\label{qv}
In this section we revisit in detail QVF as originally put forward by \cite{2008EL.....8269002D} in order to correct some misprints present there and elucidate the physical ideas involved. The energy loss to be derived basically stems from a backreaction procedure, thus approximative. It would be of interest to contrast it with the result coming from direct analyses of the field equations for a nonlinear Lagrangian ${\cal L}$ (especially the effective nonlinear Lagrangian of QED), following the lines of \cite{1955AnAp...18....1D}. We plan to do this elsewhere.

The phenomenon of QVF is basically an energy loss mechanism due to the interaction of a magnetic dipole ($\vec{m}$) with angular velocity $\vec{\omega}$ (taken to be in the $z$-direction) and the magnetization $\vec{M}_{qv}$ it produces in a surrounding medium. The associated induced magnetic field exerts a torque on the rotating magnetic dipole, leading the latter to lose energy. The infinitesimal version of such power is given by \citep{2008EL.....8269002D}
\begin{equation}
d\dot{E}_{qv}(\vec{r},t+r/c)\doteq \vec{m}(t+r/c)\times d\vec{B}_{qv}(\vec{0},t+r/c)\cdot \vec{\omega}\label{dEqvf},
\end{equation}
where $r$ is the norm of the radial vector $\vec{r}$, connecting the element of volume $dV$ (that generates the infinitesimal magnetic field $d\vec{B}_{qv}$) to the origin of the system (where the magnetic dipole is supposed to be) and 
\begin{equation}
d\vec{B}_{qv}(\vec{0},t+r/c)=\frac{3\vec{r}[d\vec{m}_{qv}(\vec{r},t)\cdot \vec{r}]}{r^5}-\frac{d\vec{m}_{qv}(\vec{r},t)}{r^3}\label{dBqv},
\end{equation}
where $d\vec{m}_{qv}\doteq \vec{M}_{qv}dV$. Notice from the above equations that retarded effects were considered and only the dipole approximation has been used for the determination of the magnetic fields. Only for completeness, recall that the magnetic field generated by the magnetic dipole is given by
\begin{equation}
\vec{B}(\vec{r},t)=\frac{3\vec{r}[\vec{m}(t-r/c)\cdot \vec{r}]}{r^5}-\frac{\vec{m}(t-r/c)}{r^3}\label{Bmagdip}.
\end{equation}
In the following we shall attempt do describe QVF related to the external region of a star of radius $R$ that generates $\vec{m}$ and also rotates with angular velocity $\vec \omega$, assuming that $\vec{m}$ makes an angle $\phi$ with its axis of rotation. In other words, kinematically,
\begin{equation}
\vec{m}(t)=m_0[\hat{z}\cos\phi + \hat{x}\sin\phi \cos(\omega t) + \hat{y}\sin\phi \sin(\omega t)]\label{mdip},
\end{equation}
where $m_0$ is given by Eq.~(\ref{m}).

Our description is only meaningful when there is a medium for $r\geq R$, since the one of the star does not contribute to QVF directly, but only to determine $\vec{m}$. As already mentioned, an effective medium does take place whenever the electromagnetism is nonlinear and its byproducts are as real as any physical medium. Its magnetization due to a nonlinear theory of the electromagnetism ${\cal L}$ is (Gaussian units) \citep{1934RSPSA.144..425B, 1975clel.book.....J}
\begin{equation}
\vec{M}_{qv}\doteq \frac{1}{4\pi}\left(\vec{B}+4\pi\frac{\partial {\cal L}}{\partial \vec{B}}\right)\label{Mqvdef}.
\end{equation}
The functional form of the Lagrangians that we shall be interested in this work is
\begin{equation}
{\cal L}= \frac{1}{16\pi}(-F + \mu F^2)\label{L},
\end{equation}
with $F\doteq F^{\mu\nu}F_{\mu\nu}= 2(B^2-E^2)$, $F_{\mu\nu}$ the electromagnetic field tensor \citep{1975ctf..book.....L} and for a given vector $\vec{X}$, $X^2\doteq \vec{X}\cdot\vec{X}$. Besides, we will assume that $|F|\ll 1/{\mu}$, which means we shall work within the weak field limit to a nonlinear theory whose scale field is proportional to $1/\sqrt{\mu}$ (see its motivation in section~\ref{int} in the scope of QED). Let us consider that $B^2\gg E^2$, which is exacty the case for extended astrophysical bodies. Then, substituting Eq.~(\ref{L}) into Eq.~(\ref{Mqvdef}) we are left with
\begin{equation}
d\vec{m}_{qv}(\vec{r},t)\doteq \vec{M}_{qv}dV=\frac{\mu}{\pi}B^2\vec{B}(\vec{r},t)dV\label{mqvL}.
\end{equation}
From Eqs. (\ref{dEqvf}), (\ref{dBqv}), (\ref{Bmagdip}), (\ref{mdip}) and (\ref{mqvL}), 
the infinitesimal mean value over a period $(2\pi/\omega)$ of the energy loss due to QVF in a given polar direction $\theta$ and radial distance from the origin $r$ is
\begin{eqnarray}
&&\frac{\langle d\dot{E}_{qv}\rangle}{\mu dV} =\frac{m_0^4 \omega  \sin ^2\phi}{128 \pi  r^{12}}\{156 \cos (2\theta)+81 \cos (4\theta)-557 \nonumber  + \\ && 3\cos (2\phi)[-84 \cos (2 \theta)+45 \cos (4\theta)-25]\}\sin\left(\frac{2\omega r}{c}\right)\label{dEqvav}.
\end{eqnarray}
One can then integrate the above equation for $r\geq R$ and all angular directions and after simple calculations obtain 
\begin{equation}
\langle\dot{E}_{qv}\rangle\simeq -\frac{24\mu m_0^4 \omega^2\sin^2\phi}{5cR^8}\label{Eqvm},
\end{equation}
assuming that $\omega R/c\ll 1$. Notice that the aforesaid integral can be solved exactly and thus further powers of $R\omega/c$ can be readily obtained whenever necessary.

We stress that the electromagnetic properties of the star can be entirely summarized by its mean surface magnetic field for the dipole approximation, Eq.~(\ref{m}).Therefore, general relativistic corrections to this classical model could all be incorporated into $B_0$. It has already been shown by \citet{2015ApJ...799...23B} that they mainly lead to the decrease of $B_0$ concerning its classical counterpart by a multiplicative factor related to the compactness of the star. Thus, classical analyses already suffice to obtain the main physical radiation aspects of pulsars.

As a realization of our analyses, let us consider now the Lagrangian density of QED. In this case \citep{2010PhR...487....1R}, 
\begin{equation}
\mu= \frac{\alpha}{90\pi B_c^2}\label{muQED},
\end{equation}
where $\alpha$ is the fine structure constant. Substituting Eq.~(\ref{m}) into Eq.~(\ref{Eqvm}) and taking into account Eq.~(\ref{muQED}), we finally have
\begin{equation}
\langle\dot{E}_{qv}^{qed}\rangle=-\frac{4\alpha}{75}\frac{B_0^4R^4\pi \sin^2\phi}{B_c^2cP^2}\label{EqvQED},
\end{equation}
where we have considered, instead of the frequency of the star, its period $P$, $\omega=2\pi/P$. Notice that this result is half of the one reported by \citet{2008EL.....8269002D}. The main reasons for that are believed to be the factor 2 within the last multiplicative sinusoidal term on the right-hand side of Eq.~(\ref{dEqvav}), obtained when dealing with retarded effects and also the definition of the surface magnetic field due to a magnetic dipole, Eq.~(\ref{m}), in terms of an area average procedure.

On the other hand, as it is well known and has already been mentioned in the previous section,
pulsars also lose energy via the magnetic dipole radiation, $\dot{E}_{d}\doteq P_{dip}$, i.e. [see Eqs. (\ref{eq:dipole}) and (\ref{m})], \citep{2001thas.book.....P, 1975ctf..book.....L}
\begin{equation}
\dot{E}_{d}= -\frac{2}{3c^3}|\ddot{\vec{m}}|^2= -\frac{16\pi^4B_0^2 R^6\sin^2\phi}{3P^4c^3}\label{Ed}.
\end{equation}
We shall surmise in this work that the total energy of the star is provided by its rotational counterpart, $E_{rot}=I\omega^2/2$, and its change is attributed to both $\langle\dot{E}_{qv}\rangle$ and $\dot{E}_{r}$. Therefore,
\begin{equation}
\dot{E}_{rot}\equiv \langle\dot{E}_{qv}\rangle +\dot{E}_{d}\label{Erotdef}.
\end{equation}
Thus, from Eqs. (\ref{eq:Edot}), (\ref{EqvQED}), (\ref{Ed}) and (\ref{Erotdef}) the evolution of the period of a star is given by
\begin{equation}
\dot{P}=  \frac{4\pi^2B_0^2R^6\sin^2\phi}{3IPc^3}+ \frac{\alpha B_0^4R^4P\sin^2\phi}{75I\pi c B_c^2}\label{Et}.
\end{equation}
From the above equation one clearly sees that its period of rotation tends to increase with time (it slows down as time goes on) and that the first term on the right-hand side is predominant for systems with small periods, the opposite being true for its second one.  Therefore, one would expect that in magnetized white dwarfs (see e.g.~\cite{2015SSRv..191..111F}), super-Chandrasekhar White Dwarfs (SChWDs) \citep{2013PhRvL.110g1102D}, Soft Gamma-Ray Repeaters (SGRs) and Anomalous X-ray Pulsars (AXPs)~(see McGill Magnetar Catalog~\citep{2014ApJS..212....6O}) the effect of QVF (more likely its generalization by means of the insertion of higher powers of $F$ in the Euler-Heisenberg Lagrangian density in order to describe supercritical magnetic fields) could be significant. This shall be investigated elsewhere.

\section{QVF braking index for constant $I$, $B_0$ and $\phi$}
\label{braking}
\begin{figure}
\includegraphics[width=\linewidth,clip]{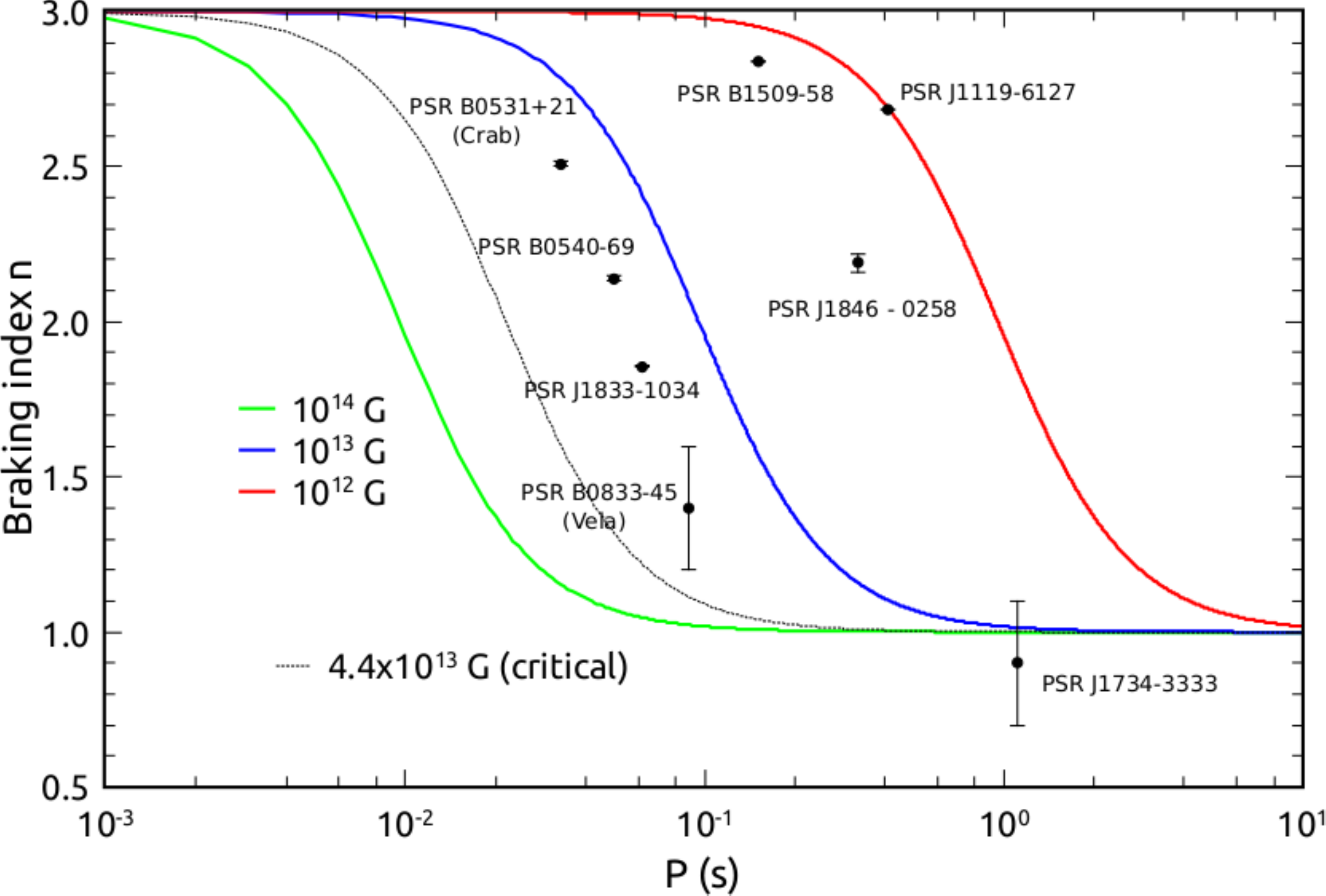} 
\caption{Braking index $n$ [see Eq.~(\ref{nEqvEd})] for some pulsars when both the classic dipole and QVF models are taken into account. }\label{fig2}
\end{figure}

Now we turn our attention to the braking indices. Typically, $n$ is associated with pulsars and it is a measure of its spin down's slope curve.
It can be used to determine how close a rotationally powered pulsar is 
from the magnetic dipole model pertaining to its energy losses, namely 3.
Among the known radio pulsars, only young pulsars have braking indices
measured accurately. We emphasize that all the reported ones have values smaller than 3, see Table~\ref{ta1} and Fig.~\ref{fig2}.

This quantity has a special relevance for compact stars, since it is a direct observable. From Eqs.~(\ref{EqvQED}), (\ref{Ed}) and (\ref{Et}), it is simple to show that for the model given by Eq.~(\ref{Erotdef}) in the case where $B_0$, $I$, $R$ and $\phi$ are all constants (physically equivalent to 
having $\dot{P}/P\gg (|\dot{B}_0|/B_0, |\dot{I}|/I$ and $|\dot{\phi}|/\tan\phi$)), $n$ is given by
\begin{equation}
n=n_0\doteq
3-\frac{2}{1+\frac{\dot{E}_{d}}{\langle\dot{E}_{qv}\rangle}}=
3-\frac{2\alpha B_0^2c^2P^2}{B_0^2c^2P^2\alpha+100\pi^3 B_c^2R^2} \label{nEqvEd}.
\end{equation}
Notice from Eq.~(\ref{nEqvEd}) that when $\dot{E}_{qv}\gg \dot{E}_r$ [$B_0\gg 5\sqrt{\pi/\alpha}B_c(R\omega/c)$], one has that $n\rightarrow 1$. When the classical radiation term is much larger than QFV the braking index tends to $3$. Since the second term on the right-hand side of the above equation is never larger than $2$, we conclude that $1<n<3$.  Besides, given a $n$ in such an interval, one shows that its corresponding $B_0$ is
\begin{equation}
B_0= \frac{5 \sqrt{(3-n)\pi} }{\sqrt{\alpha  (n-1)}}\left( \frac{R\omega}{c}\right) B_c\label{B0n}.
\end{equation}

We recall that Eq.~(\ref{nEqvEd}) is only physically relevant in the context of QVF when $B_0\ll\sqrt{90\pi/\alpha}B_c\approx 200 B_c$. From Eq.~(\ref{B0n}), this means that
\begin{equation}
\sqrt[]{\frac{3-n}{n-1}}\left(\frac{R\omega}{c}\right)\ll 1\label{nconst}.
\end{equation}
The proximity of $n$ from unit (from above) is dictated solely by the system's kinematic aspects [naturally $R\omega/c<1$]. 
For a typical pulsar, for instance, $R\omega/c \approx 10^{-3}$ and one sees that any $n \gtrsim 1.001$ leads the above inequality to be fulfilled. This means that within QVF surface magnetic fields for pulsars with $1<n<3$ can be at most of the order of the critical magnetic field of QED. Figure \ref{fig2} shows this is exactly the case for all associated pulsars 
(the case $n>3$ is clearly not contemplated in the simple model analyzed here and shall be investigated in the next section; as one physically expects, subcritical magnetic fields will also raise there but non-null $\dot{B}_0$ or $\dot{\phi}$ will be required). Notice that the pulsar PSR J1734-3333 seems to be a very special case. It has a braking index of $n=0.9\pm 0.2$ (see \cite{2011ApJ...741L..13E} for details). This value is well below 3, and in light of our analyses it indicates that QVF could be the most relevant mechanism for the energy loss of the system.

As in the case of the classic magnetic dipole radiation model, one can solve Eq.~(\ref{Et}) for $B_0$ when assuming that all the other quantities are given, and such result is
\begin{equation}
B_0^2=  \frac{50\pi^3B_c^2R^2}{c^2P^2\alpha}\left[\sqrt[]{1+\frac{3\dot P I P^3c^5\alpha}{100B_c^2 R^8\pi^5\sin^2\phi}}-1 \right]\label{B02}.
\end{equation}
The consistency of QVF with observational quantities demands that the averaged surface magnetic fields given by Eqs. (\ref{B02}) and (\ref{B0n}) agree. This can be done by fixing some of the free parameters of the model. One of the most primitive ones in this regard is the angle a magnetic dipole makes with the axis of rotation of the star. Thus, from the aforesaid equations, one shows that
\begin{equation}
\sin^2\phi = \frac{3c^5\dot{P}I (n-1)^2P^3\alpha}{800\pi^5B_c^2(3-n)R^8}\label{sin}.
\end{equation}
An immediate outgrowth of the above equation is

\begin{equation}
\frac{M}{R^6} < \frac{2000 \pi^5 B_c^2}{3 c^5\alpha P^3 \dot{P}}\frac{(3-n)}{\;(n-1)^2},\label{MR6}
\end{equation}
where we assumed that $I = 2MR^2/5$, i.e., the moment of inertia of a homogeneous sphere. Care should be taken here concerning the physical interpretation of Eq.~(\ref{MR6}). It is not a necessary constraint that the mass and the radius of a pulsar must satisfy. It is solely a byproduct of the assumptions underlying Eq.~(\ref{nEqvEd}). 
Table \ref{ta1} shows the $\phi$'s associated with Eq.~ (\ref{sin}) for all pulsars with known braking indices. For the cases where they cannot be found, it is simple to conclude that the associated changes needed to be done in the fiducial parameters are unrealistic. 
Indeed, using Eq.~(\ref{MR6}) for $M=1.4 M_\odot$, the radii of PSR J1846-0258, PSR J1119-6127 and PSR B1509-58, would have to be larger than 29 km, 38 km, and  29 km, respectively, which are utterly improbable for pulsars. This implies that one should actually take into account the evolution of other parameters into the braking index, such as the ones ignored to obtain Eq.~(\ref{sin}). We shall come back to this issue in the next section.

\textcolor{white}{abcd}
\begin{table*}
\caption{Estimates of $\phi$ for pulsars with known braking index.}
\begin{ruledtabular}
{\begin{tabular}{@{}cccccc@{}} 
Pulsar & $P$~(s) &$\dot{P}~(10^{-13}$~s/s) &n & Ref. &$\phi$  \\ \hline
PSR B0833-45 (Vela) &0.089&  1.25& $1.4\pm0.2$  &~\cite{1996Natur.381..497L} &$\sim6.0^{\circ}$\\
PSR B0540-69 &0.050 &4.79 &$2.140\pm0.009$ 
&~\cite{2007ApSS.308..317L} &$\sim19.6^{\circ}$\\
PSR J1846-0258 &0.324 &71 &$2.19\pm0.03$ &~\cite{2015ApJ...810...67A} &- \\
PSR B0531+21 (Crab) &0.033 &4.21 &$2.51\pm0.01$  &~\cite{1993MNRAS.265.1003L} &$\sim17.3^{\circ}$ \\
PSR J1119-6127&0.408&40.2&$2.684\pm0.002$ &~\cite{2011MNRAS.411.1917W}  &-  \\
PSR B1509-58   &0.151&15.3 &$2.839\pm0.001$  &~\cite{2007ApSS.308..317L} &-\\ 
PSR J1833-1034 &0.062 &2.02 &$1.8569\pm0.0006$ &~\cite{2012MNRAS.424.2213R} & $\sim 11.3^{\circ}$ \\
PSR J1734-3333 &1.17&22.8&$0.9\pm0.2$~$^*$  &~\cite{2011ApJ...741L..13E} &$ \sim28.6^{\circ}$ \\
PSR J1640-4631 &0.207&9.72&$3.15\pm0.03$  &~\cite{2016ApJ...819L..16A} &- \\
\end{tabular} \label{ta1}}
\end{ruledtabular}
\\
{$^*$ We adopted n=1.01 to calculate $\phi$. }
\end{table*}

\section{Pulsar's evolutionary aspects within the scope of QVF}
\label{mag_evol}

It is very likely that pulsars, due to their dynamic nature, should always present important temporal changes in quantities other than $P$. This signifies that Eq. (\ref{sin}) possibly does not represent the inclination of the magnetic moments of realistic pulsars [which is equivalent to saying that Eq. (\ref{nEqvEd}) is not the most adequate equation for the braking index]. Therefore, more complex scenarios should be investigated, generalizing the results of the previous section.

Let us start with the situation in which both $\phi$ and $B_0$ are time-dependent. The case $I=I(t)$ seems unrealistic for the isolated pulsars we are investigating, or, at least, less relevant than the time dependence of $B_0$ and $\phi$. From Eqs.~(\ref{n}) and (\ref{Et}), one can readily show in this case that 
\begin{equation}
n=n_0-2\frac{P}{\dot P}\left[\frac{\left(5-n_0\right)}{2}\frac{\dot B_0}{B_0}+\dot{\phi}\cot{\phi}  \right] \label{nBphi},
\end{equation}
where $n_0$ is the braking index for the case both $B_0$ and $\phi$ are constants, Eq. (\ref{nEqvEd}), and $B_0$ will be assumed to be given by Eq. (\ref{B02}). [In this case, as self-consistency naturally demands, $\dot{\phi}$ will be the same as coming from Eq. (\ref{nBphi}), given $\dot{B}_0$ and $n$, or direct analyses of Eq.~(\ref{B02}).]

It is believed that magnetic fields should decay in pulsars [usually due to the Ohmic decay, Hall drift and ambipolar diffusion \citep{1988MNRAS.233..875J,1992ApJ...395..250G}] on timescales of order $(10^{6}-10^{7})$ yr \citep[see e.g.,][and references therein]{1992ApJ...395..250G, 2015MNRAS.453..671G}. [Nevertheless, there are also suggestions that the timescales for $B_0$ could actually be smaller, of order $10^5$ yr \citep{2014MNRAS.444.1066I,2015AN....336..831I}.] Thus, bearing in mind that magnetic fields in the context of QVF for pulsars are of the order of $(10^{12}-10^{13})$ G [see Fig. \ref{fig2}], let us assume in what follows $\dot{B}_0<0$ and $|\dot{B}_0|$ of order $(10^{-2}-10^{-1})$~G/s. (This is estimated directly from the above-mentioned usual timescales $T_B$ such that $|\dot{B}_0|\sim B_0/T_B$.) Our analyses for this case concerning $\dot \phi$, taking into account the braking indices of the pulsars under interest, are summarized in Table \ref{ta2} for the representative angle $\phi = \pi/4$ [see Eq. (\ref{nBphi})]. (Since the physically relevant values of $\dot{B}_0$ are small, the conclusions that ensue are essentially the same as the case $B_0$ constant.) Notice that some pulsars have positive $\dot \phi$'s, while others have negative ones, and all of them present subcritical magnetic fields (thus clearly showing the self-consistency of QVF also in this more complex scenario). Special attention should be paid to the Crab pulsar. The value $\dot \phi \simeq 3 \times 10^{-12}$ rad s$^{-1}$ has been observationally inferred to it \citep{2013Sci...342..598L, 2015MNRAS.446..857L,2015MNRAS.454.3674Y}, which has the same sign and magnitude as that one predicted by QVF, and thus could always be related to a specific angle $\phi$ there. We emphasize that the same analyses as the ones above could be done in the scope of the classic magnetic dipole model. In this case one can easily verify that all pulsars in Table $\ref{ta1}$ with $n<3$ are such that $\dot \phi>0$ and it is of order $10^{-12}$ rad/s [see Eq. (\ref{nBphi}) for the formal case $\alpha \rightarrow 0$].
Therefore, measurements of $\dot \phi$ for other pulsars (specially the ones that present $\dot{\phi}<0$ in the context of QVF) could easily falsify any of these models for given mechanisms of magnetic field decay and evidence their underlying physics [e.g. neutron star precessions could lead to $\dot{\phi}>0$ \citep{2015MNRAS.451..695Z,2016MNRAS.455.1845K}]. For the special pulsar PSR J1640-4631, both QVF and the classic magnetic dipole model result in $\dot{\phi}<0$, but the former model predicts a faster rate of change than the latter. Finally notice that all $\dot{\phi}$'s in Table \ref{ta2} are positive only when $\dot{B_0} \lesssim -10^2$ G/s, always leading to values larger than their classical counterparts. The difficulty in this case, though, would be the physical explanation of timescales at least three orders of magnitude smaller than the ones coming from known mechanisms of magnetic field decay.
%
\begin{table}
\caption{Estimates of $\dot\phi$ for the 
same pulsars as in Table \ref{ta1}, with
$\dot{B}_0=~-0.05$~G/s, for the representative inclination angle $\phi=\pi/4$.}
\begin{ruledtabular}
{\begin{tabular}{@{}ccc@{}} 
Pulsar&$\dot\phi$~($10^{-12}$~rad/s)& $B_0$ ($10^{12}$~G) \\ \hline
PSR B0833-45 (Vela) &$ 0.8$ &6.2\\
PSR B0540-69 &$ 2.3$ &9.3 \\
PSR J1846-0258 &$ -12$ &17 \\
PSR B0531+21 (Crab) &$2.3$ &7.7 \\
PSR J1119-6127&$ -8$ &14 \\
PSR B1509-58 &$ -7.5$ &14\\
PSR J1833-1034 &$ 1.3$ &6.9\\
PSR J1734-3333 &$  0.02^*$ &9.5 \\
PSR J1640-4631 &-4.4&11\\
\end{tabular} \label{ta2}}
\end{ruledtabular}
\\ {$^*$ We have adopted n=1 here. }
\end{table}

\begin{figure}
\leavevmode
\centering
\includegraphics[width=\linewidth,clip]{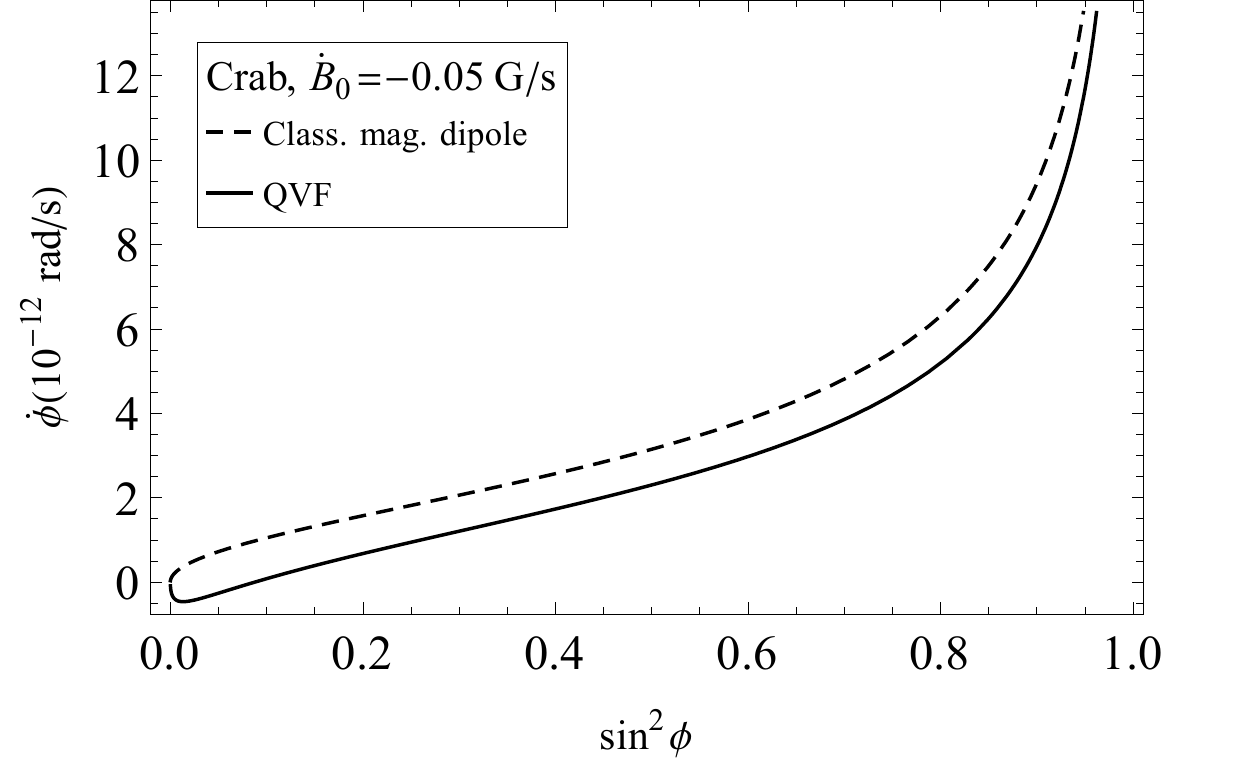}
\caption{{Instantaneous evolution of $\dot \phi$ as a function of the inclination angle $\phi$ for the Crab pulsar [see Eq. (\ref{nBphi}) and Table \ref{ta1}] with $\dot{B}_0= -0.05$G/s for both QVF and magnetic dipole ($\alpha \rightarrow 0$) models. Notice that for $\phi\lesssim 5^{\circ}$ QVF analyses are not reliable because we are close to the threshold of its validity [see Fig. \ref{figB0_Crab}]. 
For the parameters in this figure, one sees, for instance, that $\dot\phi \simeq 3\times 10^{-12}$rad/s (Crab's inferred inclination rate) would imply $\phi \approx 45^{\circ}$ for the classic magnetic dipole model, while $\dot\phi$ in the context of QVF would be approximately $51^{\circ}$. }}\label{figdotphi_Crab}
\end{figure}

\begin{figure}
\leavevmode
\centering
\includegraphics[width=\linewidth,clip]{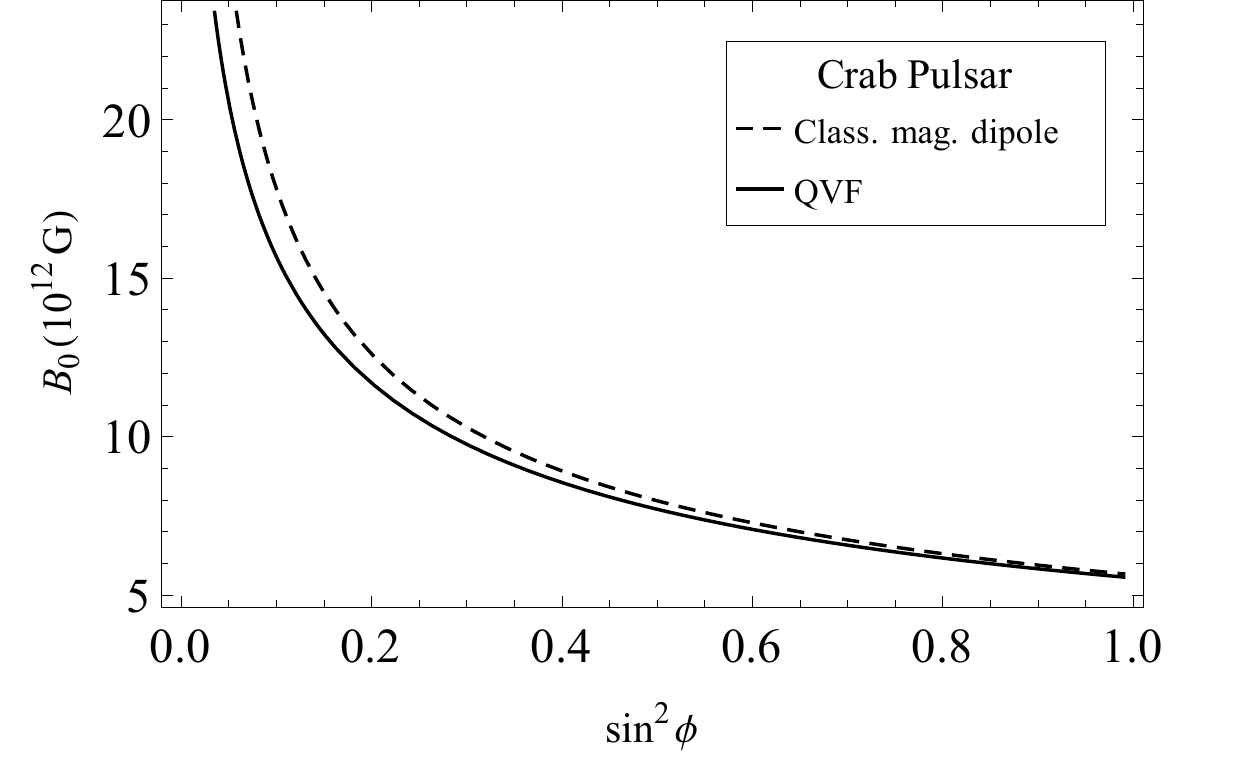}
\caption{{Instantaneous surface magnetic field a function of the inclination angle $\phi$ for the Crab pulsar [see Eq. (\ref{nBphi}) and Table \ref{ta1}] in the scope of QVF and the classic dipolar model. Notice that fields of the order of the critical one raise for $\phi\lesssim 5^{\circ}$ [see Eq. (\ref{B02})], indicating the limit of validity of QVF analyses. Besides, QVF always leads to smaller magnetic fields when compared to the classical model.}}\label{figB0_Crab}
\end{figure}

Since the Crab pulsar has an observationally inferred $\dot \phi$, let us study more precisely the implications of QVF and the classic magnetic dipole for it. Figure \ref{figdotphi_Crab} depicts the behavior of $\dot \phi$ [see Eq. (\ref{nBphi})] for the Crab pulsar in the case $\dot{B}_0= -0.05$ G/s for both above-mentioned models. Notice that for angles $\phi \lesssim 5^{\circ}$ QVF analyses are not trustworthy, since we are approaching its threshold of validity [see Fig. \ref{figB0_Crab} and Eq. (\ref{B02}) for this case]. Besides, $\dot \phi$'s related to the classic dipolar model are always larger than the ones coming from QVF, which means that in the latter model, for a given $\dot{\phi}$, the actual (instantaneous) $\phi$ is always larger than the one coming from the former model. For instance, for $\dot{\phi}\simeq 3\times 10^{-12}$ rad/s, the classic dipole model implies $\phi \approx 45^{\circ}$, while QVF predicts $\phi\approx 51^{\circ}$. Only for completeness, in Fig. \ref{figB0_Crab} we plot the instantaneous surface magnetic field for the Crab pulsar as a function of $\phi$. It is evident there that for angles larger than $\phi \gtrsim 15^{\circ}$, only subcritical magnetic fields raise. In Fig. \ref{figdotphi_Crab} one can also see that for the QVF model there is a nontrivial angle such that $\dot \phi =0$. This is already expected due to the existence of a $\phi$ satisfying the simpler case given by Eq. (\ref{sin}) [see Table \ref{ta1}]. Their proximity is simply due to the smallness of $\dot{B}_0$.  
Finally, in the case of the Crab pulsar, QVF can only be differentiated from the classic magnetic dipole model if precise measurements of its $\phi$ are available, still not the case. 

Consider now the case in which only $B_0$ is allowed to change with time. This would be a natural consequence of the existence of equilibrium angles to the directions of pulsars' magnetic dipoles. One can verify with the use of Eq. (\ref{nBphi}) that (self-consistent) $\phi$'s could only be found to the pulsars with $n<3$ that do not satisfy Eq.~(\ref{sin}) in Table \ref{ta1} when $\dot{B}_0<0$ and $|\dot{B}_0|\sim  (10^2-10^3)$ G/s. Hence, when QVF is taken into account for these pulsars, it would lead surface magnetic fields to \textit{decrease} with time and in a such a way that their rotational energy always overwhelms their magnetic one. From the previous results one also obtains that the characteristic timescales for the surface magnetic fields of the pulsars under discussion are $(10^3-10^4)$ years. 
It is interesting to note that such timescales are in agreement with the ones coming from $P/\dot{P}$ for the same pulsars. 
This suggests that the evolution of $B_0$'s in pulsars with known braking indices and not associated $\phi$ in Table \ref{ta1} should be connected with their spindown, pointing to the relevance of the mechanisms where this takes place, such as in Ruderman's neutron vortices (that will drag along protons and thus also influence the magnetic field of a pulsar) \citep{1970Natur.225..619R, 1972ARA&A..10..427R}.
We underline that the above timescales for surface magnetic fields obtained within QVF clearly contrast with the ones related to a purely magnetic dipole radiation model for pulsars, in which surface magnetic fields should \textit{increase} with time, having timescales of $(10^2-10^3)$yr \citep{1996ApJ...458..347M}. The fact that the magnetic timescales found within the scope of QVF for $n<3$ are much smaller than those ones coming from Ohmic decay and Hall drift, for instance, suggests that the associated pulsars are currently experiencing transient periods. This would be supported by PSR J1846-0258, which had $n=2.65$ six years ago \cite{2015ApJ...810...67A}. Another natural conclusion would be that the assumption of having constant $\phi$ is incorrect, as suggested by \cite{2015MNRAS.454.3674Y}. Only further observations could settle this ambiguity. Just for completeness, for the pulsars in Table \ref{ta1} that already have associated angles, one can check that QVF leads to positive $\dot{B}_0$ and of the order of $10$~G/s whenever the chosen $\phi$'s are larger than the ones satisfying Eq.~(\ref{sin}). Due to a simple continuity argument, $\dot{B}_0<0$ when they are smaller. For the former case, one sees that the associated timescales of magnetic field growth are of the order of $10^4$ yr, larger than their classical counterparts. In summary, measurements of $\dot{B}_0$ for $n<3$ could also easily falsify QVF, since it leads to both positive and negative values of such a quantity, not the case for the classic magnetic dipole model. Besides, in the case magnetic fields increase, the above mentioned models predict very different timescales for them. The same ensues for the case $n>3$, where now in both models magnetic fields should decrease with time.

\section{Discussion and Conclusions}
\label{discussion}
Since the stars we analyzed are rotating, physically relevant quantities should be time averaged (per period of rotation), as it was done in section \ref{qv}. This is specially the case for the resultant mean torque per cycle $\langle \vec{N}\rangle$ on the star's surface due to the whole effective magnetized medium surrounding it. As it is evident from the symmetry of the problem, $\langle \vec{N}\rangle$ must be collinear with the rotation axis of the star. The simplest manner to obtain it is from Eq. (\ref{EqvQED}) and the definition of the power associated with any torque ($\vec{N}\cdot\vec{\omega}$), that leads to
\begin{equation}
\langle\vec{N}\rangle= -\frac{\alpha}{75}\frac{B_0^4R^4\sin^2\phi}{B_c^2\pi c}\vec{\omega}\label{torque}.
\end{equation}
(One can also obtain $\langle\vec{N}\rangle$ as above by a direct computation, starting from the definition of the infinitesimal torque related to Eq. (\ref{dEqvf}) and then following the same procedure that led us to Eq.~(\ref{EqvQED}).) One sees from Eq.~(\ref{torque}) that the resultant time-averaged torque is anti-parallel to $\vec{\omega}$, intrinsically associated with a force proportional to the negative of the velocity. Thus, QVF leads the decrease of rotational energy of the star to be converted into heat. This explains the energy balance related to QVF. It clearly contrasts with magnetic dipole radiation in which the star's slowdown is due the emission of electromagnetic radiation. The by-products of this heat are beyond the scope of this work and shall be investigated  elsewhere.

It should be stressed that QVF is an intrinsically quantum effect related to the backreaction of the vacuum polarization on a classical magnetic dipole, leading the star to lose energy by means of a torque. 
There is no reason for it to be much smaller than the radiation associated with a classic magnetic dipole simply because the effects do not have the same nature. Only corrections to the classical magnetic dipole radiation due to the nonlinearites of the Lagrangian density are automatically small within the QVF model, see Eq.~(\ref{L}), and exactly due to that they were disregarded in our analyses. 

Let us quickly discuss some evolutionary aspects of the braking indices of pulsars in light of QVF. Since the energy loss due to QVF decreases with $P^{2}$, while the classic magnetic dipole radiation decreases with $P^4$, see Eqs. (\ref{EqvQED}) and (\ref{Ed}), QVF should be predominant only at later evolutionary times of a pulsar,
making its braking index tend to unit if $B_0$ and $\phi$ are asymptotically stationary [see Eq.~(\ref{nBphi})].
In this regard, one could tentatively state that this could be the case of [or supported by] the pulsar PSR J1734-3333, be due to its relatively large value of $P$, be to its measured braking index. At the same time, the aforesaid pulsar could also be an example that falsifies QVF in the case studied in section \ref{braking}. This naturally motivates further studies concerning PSR J1734-3333, in order to decrease the uncertainty present in its braking index [the same can be said to the pulsar PSR J0537-6910, whose normally associated braking index of $-1.5$ (see e.g., \cite{2015MNRAS.452..845H}) is not at all accurate, due to the large dispersion in $\ddot{P}$ it presents \cite{2006ApJ...652.1531M}], as well as to restrict evolutionary aspects of its $B_0$ and $\phi$.  
Whenever $\dot{B}_0$ and $\dot{\phi}$ are not asymptotically stationary, one can clearly see from Eq. (\ref{nBphi}) that several scenarios raise within QVF, even the one in which $n<1$, that can obtained when $\dot{B}_0/B_0 \geq -2\dot{\phi}/[(5-n)\tan{\phi}]$. [Here, like what happens in the classic model, one notices that when $\dot{B}_0<0$, $\dot{\phi}$ must be positive for $0<\phi<\pi/2$ and negative for $\phi/2<\phi<\pi$, indicating, thus, that $\phi=\pi/2$ is an attractor to the magnetic dipole direction. ] Another example would be the one studied in section \ref{mag_evol} in which for the time being $1<n<3$. In such a case, QVF could only constrain the temporal evolution of some pulsars' parameters (as we have done for $\dot \phi$ and $\dot{B}_0$)
and only their measurement could rule it out. Thus, generically speaking, observations alone on the braking index cannot fully falsify QFV, but only constrain it.

Quantum vacuum friction can in principle be easily distinguished from the classic magnetic dipole radiation. As we have showed in section \ref{mag_evol}, it predicts in such a scenario that $\phi$ should change with time, being either negative or positive for different pulsars, quite differently from the classic magnetic dipole model. Similar conclusions can be drawn for the evolution of $B_0$. For the Crab pulsar, both above mentioned models lead to current inclination angles that differ from each other by some degrees. This motivates further analyses in this direction.

We point out that a simplified model has been chosen in order to assess the relevance of QVF more transparently.  Actually, it is known that a pulsar should have a plasma atmosphere
\citep[see e.g.,][and references therein]{1974ApJ...192..713M} and it is not simply surrounded by vacuum. Besides modifying the standard magnetic dipole radiation model (with an extra torque), it is also expected to influence QVF due to the following reason. This plasma region would influence the resultant magnetic field felt in the (outer) vacuous region, which would directly influence the vacuum magnetization [see Eq.~(\ref{mqvL})]. This in turn would lead the quantum vacuum to exert a different torque on the star, which would change its slowdown. Clearly this is a more elaborated scenario that we shall discuss more precisely elsewhere. 

Summing up, in this work we have also taken QVF as a fundamental energy loss mechanism and we have tried to assess its relevance into the description of pulsars. In its simplest form, it leads magnetic fields to automatically be subcritical (in plain contrast, for instance, with high-B pulsars in the context of the classic magnetic dipole).
In addition, measurements of quantities other than $P$ and its derivatives for pulsars (such as $\dot{\phi}$ and $\dot{B}_0$) could easily falsify QVF for the case $n<3$. Finally, it seems that QVF should be a relevant source of energy loss for the pulsar PSR J1734-3333.

\begin{acknowledgments}
J.G.C. acknowledges the support of FAPESP (2013/15088-0 and 2013/26258-4). J.P.P. is likewise grateful to CNPq- Conselho Nacional de Desenvolvimento Cient\'ifico e Tecnol\'ogico of the Brazilian government within the postdoctoral program ``Science without Borders''. J.C.N.A thanks FAPESP (2013/26258-4) and CNPq (308983/2013-0) for partial support. We thank the anonymous Referee for valuable comments and suggestions.
\end{acknowledgments}




\bibliography{ref_qvf}
\bibliographystyle{apj}

\end{document}